# High-Speed Network Traffic Management Analysis and Optimization: Models and Methods

V.Zaborovski (vlad@stu.neva.ru)
State Technical University of Saint-Petersburg, Russia

Y.Podgurski (podg@stu.neva.ru)
Institute of Robotics and Technical Cybernetics, Russia

S.Yegorov (ye@stu.neva.ru)
State Technical University of Saint-Petersburg, Russia

Y.Shemanin (yuri@neva.ru)
Institute of Robotics and Technical Cybernetics, Russia

*Abstract.*
The main steps of automatic control methodology include the hierarchical representation of traffic management system and the formal definitions of input variables, object and goal of control of each management level. It is noted that the current set of traffic parameters recommended by ATM-forum is not enough to synthesize optimal control system. The feature of traffic self-similarity can be used to effectively solve optimal control task. The examples of synthesis of the optimal traffic control mechanism and the dynamic resource allocation algorithm are presented.

*Contents*

*Introduction*. Effective traffic management methods and algorithms development becomes the very important task due to the explosive growth of the TCP/IP Internet and multimedia applications.
Currently, they started using the methods of optimized control [1] and stochastic dynamics [2]. From system analysis point of view high-speed computer network can be represented as complicated distributed hierarchical dynamics object which parameters and structures are time dependent. This feature considerably complicates tasks of control and providing quality of service (QoS) needed. In contrast to phone and TV systems, computer systems use sophisticated protocols for access to resources. The feature of those protocols is their interaction with distributed data to perform congestion control and recovery algorithms. These algorithms give rise to a variety of interesting dynamics [3, 4]. That feature of computer communications allows using of facilities and methods of intelligent control for adaptation of applications to used network resources. Such shift towards intelligent control creates basis for integration of existing communication systems into united Intelligent Information Infrastructure (III) of computer telecommunications. In this paper, our interest is in analyzing the models and methods which can be used for traffic management algorithms based on optimization and adaptation procedures.

The most difficult aspect of high-speed network development and hotly debated topic is management design. The main goal in designing is to maintain the QoS while attempting to make maximal use of network resources. To define the requirements to network control a network management system may be decomposed into three parts: congestion control techniques, traffic management functions and information access/search algorithms. The selection depends upon the duration and severity of management. Congestion control deals only with problem of reducing load during overload. Traffic management functions have to be intelligent procedures in their nature since they try to provide needed QoS. For this purpose the network continuously monitors its traffic, provides real-time data mining and feedback to the source end systems. The amount of information available in networks is growing very fast. In order to decrease user load the intelligent solutions with data mining algorithms may be used as a part of network information access/search algorithms.

New management requirements of future network infrastructure can be made basing upon the analysis of application tasks that use the new abilities new networking technologies have.

*Traffic management: a formal approach.*
At high speed the connection holding times become shorter and bursty. Therefore high-speed networks to succeed, they should be able to handle the burst traffic efficiently basing on sophisticated management functions.

A number of different traffic control mechanisms have been proposed in the literature. Mostly the verbal (not formal) definitions of management functions are given and no numerical characteristics of management efficiency have been done.

For comparative analysis and complex management performing based on different control schemes there is a need for the uniform generic formal theoretical approach.

With few exception the wide range of control mechanisms have been described may be classified in terms of automatic control theory.

We propose to use existing Automatic Control and Operational Research approaches to describe, analyze and design the network management strategy and different level's traffic control mechanisms.

Proposed approach implies:
- traffic management representation as a hierarchical system with different levels of management and
- the formal definition of each management level.

For many traffic management tasks, such a hierarchical presentation is natural and even unavoidable. For example, a complete traffic management strategy should include several congestion control and avoidance schemes that work at different protocol levels, and that can handle congestion of varying duration [12]. From the automatic control point of view, it means that there is a set of automatic control systems that have different control rates and respond to the input force changes in appropriate time intervals (short or long). Despite the differences in management goals and time scale of levels, we propose to give the formal definition of every management level and to formalize the task of control synthesis based on:

- the model of input variables,
- the model of object under control,
- the formal definition of efficiency criterion,
- the methods of the optimal regulator deriving,
- the set of variables (parameters) transmitted to other levels of management.

This approach promises
— to facilitate the design and comparing different control schemes by using the automatic control theory methods and taking into account the characteristics of feedback and open-loop regulators, such as: stability, precision, controllability, control latency, complexity;
— to facilitate the technology partnership and hens to accelerate the standardization of technology ideas;
— to define a formal task of creating optimal and/or adaptive control mechanisms.

This leads us to the discussion: what does efficient management mean? What are the numerical metrics of management efficiency in general and of each control mechanism in particular? What are the adequate patterns (mathematical models) of object under control and of input forces for different levels of management? Below, without loss of generality, we discuss our considerations of choosing models and efficiency criteria oriented to high-speed network with cell-based traffic (ATM-technology).

*ATM Forum specifications and individual traffic source model.*
A primary role of traffic management in ATM networks is to protect the network and the end-system from congestion in order to achieve network performance objectives and use network resources efficiently. The ATM Forum has defined a set of main traffic management functions that may be used in appropriate combinations depending on the service category [5].

Note that the input variables for these functions can be divided into individual source traffic (for example, for Usage Parameter Control) and total network workload (for example, for ABR Flow Control). In order to take into account data stream characteristics while synthesizing the control law, we have to develop models of individual traffic source (IND-model) as well as multiplexed traffic source (MIX-model). The robust model has to be built basing upon actual traffic parameters. In accordance with ATM Forum specifications [5], each actual connection is specified by traffic contract which includes the negotiated values of traffic parameters - Peak Cell Rate (**PCR**), Sustainable Cell Rate (**SCR**), Maximum Burst Size (**MBS**) and QoS parameters - Cell Loss Ratio (**CLR**), Cell Transport Delay (**CTD**) and Cell Delay Variation (**CDV**).

That is why the ability to handle those parameters to create different types of control schemes is very important. First of all the adequate model of the traffic source is needed.

The individual source traffic can be defined as a discrete parameter stochastic process $\{X_n\}$ representing the spacing of cell flow. The process has to satisfy the restrictions derived from the traffic parameters: Mean $\{X_n\}=1/SCR$, $X_n \geq 1/PCR$, maximum length of continuous sequence of minimal values **($X_n$=1/PCR)** must be not greater then **MBS**. The analytical expression of a such cell flow is absent.

As a first step of developing unified formal IND-model we propose a Petri net based model of individual traffic source, which takes into account the real traffic parameters and may be used for both simulating and some analytical investigations. Petri net is a well-known formal tool for modeling the behavior of discrete systems and have an analytical and graph interpretation [6,7].

We use the notion of "synchronic distance" [8] to synthesize and analyze Petri net models.

The synchronic distance constitutes an important system invariance property relating to the synchronization of two events (the firing of two transitions). It reflects the degree of independence between two transitions or more precisely, the degree to which the two transitions may deviate from the mean firing count ratio as specified by the respective components of reproduction vector of Petri net.

The synchronic distance may be formally defined as follows [8]. Let m is a live marking and **P(m)** is the set of possible firing sequences of the Petri Net PN. Let **r** be a reproduction vector of PN whose components $r_a$ and $r_b$ associated with the transitions $t_a$ and $t_a$, respectively, are both greater then zero. Then

$$syn(t_a,t_b)=\max_{p \in P(m)} (1/d)*(|r_b*h_a(p)-r_a*h_b(p)|)$$

is said to be the weighted synchronic distance between $t_a$ and $t_b$, where $h_a(p)$ and $h_b(p)$ are the components of the transition vector **h(p)** which specify how many times $t_a$ and $t_b$ respectively, fire during the process **p** and **d** is the greatest common divisor of $r_a$ and $r_b$.

The synchronic distance fits very well to be a measure of maximum burst size in the source model based on Petri nets.

The traffic source model is defined as Petri net (Fig.1).

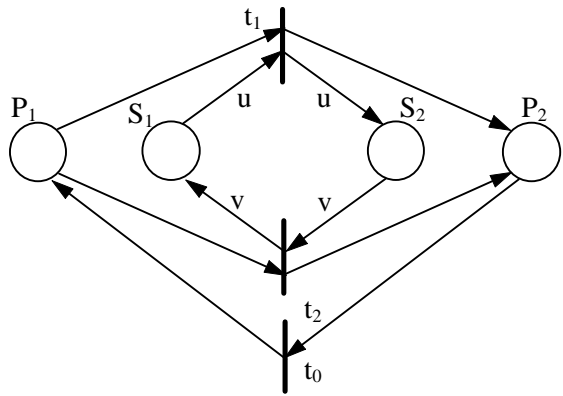

Fig. 1

The arcs $(s_1,t_1)$ and $(t_1,s_2)$ of the model both have weight **u**, and the arcs $(s_2,t_2)$ and $(t_2,s_1)$ both have weight **v**. Hence, whenever $t_1$ fires, **u** tokens move from $s_1$ to $s_2$, and whenever $t_2$ fires, **v** tokens move from $s_2$ to $s_1$. The initial marking $M_0$ of Petri net is
**$M_0=(M(p_1)=1, M(s_1)=z \geq u+v-1, M(s_2)=0, M(p_2)=0)$**.

The values **$M(s_1)$, u, v** are the functions of service category and traffic parameters.

The transition $t_0$ may be considered as an internal clock of the source and has a finite firing time $T_0$. The interval between two consecutive occurrences of $t_0$ then is taken to be a time slot of internal clock. Transitions $t_1$ and $t_2$ fire instantly after enabling. As $t_1$ and $t_2$ may fire only alternately with $t_0$ if $M(p_1)+M(p_2)=1$ the firing of $t_1$ may be viewed as emulating the sending of a cell whereas the firing of $t_2$ emulates the time slots during which no sending takes place.

It may be shown [8] that model regulates, for reproducing processes **Pr**, the firing count ratio between $t_1$ and $t_2$ as $h_1(Pr)/h_2(Pr)=v/u$, where $h_i(Pr)$ is the component of reproduction vector $h(Pr)$ associated with the transition $t_i$. The Petri net model is obviously live if $z=M(s_1) \geq u+v-1$. The actual number of tokens **z** which constitutes a live marking of model determines how rigorously the firing count ratio $h_1(Pr)/h_2(Pr)=v/u$ is enforced. This may be illustrated by mentally executing the Petri net with stepwise increasing the initial number of tokens in place $s_1$. We may observe that we introduce an increasing degree of freedom as to the choice of either firing $t_1$ or $t_2$ under a particular token distribution. A quantitative measure for this phenomenon is a synchronic distance formally defined above. It may be shown [8] that for Petri net of Fig.1 and a live marking with $M(s_1)=z-q$, $M(s_2)=q$ and $M(p_1)+M(p_2)=1$, the synchronic distance between $t_1$ and $t_2$ is given as

$Syn(t_1,t_2) = \lceil (z-q)/d \rceil + \lceil q/d \rceil$,

where $\lceil x \rceil$ denotes the greatest integer $\leq x$ and **d** is the greatest common divisor of $h_1(Pr)$ and $h_2(Pr)$. The different classes of traffic source may be efficiently modeled by using the proposed model [10, 12].

For example, consider the **VBR** class traffic source model. The relations between the Petri net characteristics and traffic source parameters are as follows:

$$M(s_1)=MBS*(PCR-SCR)/w,$$

where **w** is the greatest common divisor of **SCR** and (**PCR**-**SCR**),

$$u=M(s_1)/MBS, \qquad (1)$$

$$v=M(s_1)*SCR/[MBS*(PCR-SCR)],$$

$$T_0=1/PCR.$$

Then the firing sequence of transition $t_1$ will exactly correspond to the **VBR** traffic with
$PCR=1/T_0$, $SCR=1/T_0*v/(u+v)$, $MBS=M(s_1)/u=Syn(t_1,t_2)/u$,

The above traffic source models are based on the "top" values of traffic parameters and can be used for simple calculation of network resources distribution for a worst case.

*An example of optimal resources allocation.*
Let us consider the overall access control and resource assignment scheme in order to investigate the resource allocation problem in the ATM network. It is taken from [1] and depicted in Fig.2

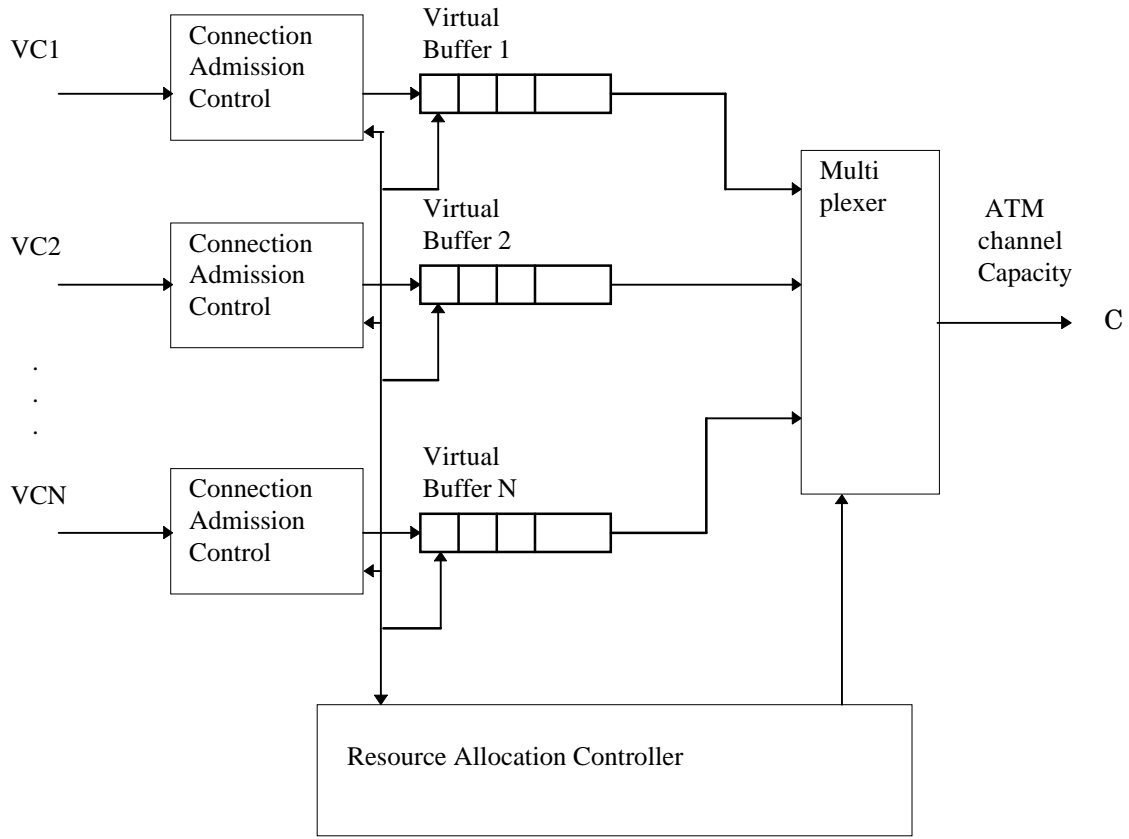

Fig.2

It includes:

N virtual channels (VC) ;
N virtual admission controllers;
N virtual buffers;
the Multiplexer and
Resource allocation controller.

Each admission controller implements a decision rule for the acceptance of incoming calls. The resource allocation controller "sees" the traffic and QoS parameters of admitted connections and computes the bandwidth allocation and buffer size for each **VC**. It divides the total capacity **C** (bandwidth in Cells/s) of ATM channel) into virtual capacities $C_i$ and defines the virtual channel buffer size $S_i$ to provide the minimal cell loss. The total available buffer size is **S**.

We suppose that:

1. The traffic parameters of each **VC** are given in terms of **PCR**, **SCR** and **MBS** .
2. $\sum SCR_i \leq C$ .
3. The QoS parameters of each **VC** are given in terms of **CLR** and **CDV**. We assume that **CTD** is mostly caused by propagation delay and not depends on resource allocation decision.
4. Each **VC** remains within the traffic contract during the connection lifetime.

The problem that needs to be solved is to develop a resource allocation scheme which honors the real traffic contract (traffic and QoS parameters) of the active **VC** and provides the efficient use of network resources.

*The model of input variables.*

Let us first discuss a case when virtual bandwidth $C_i$ for each channel is assigned on the basis of **SCR** of $VC_i$ (**PCR** for **CBR** class connection).

That creates the task of the determining of minimum size of virtual buffers that provides no cell-loss in virtual circuits. That task can be solved separately for every circuit using simplified Petri net model of resource assignment scheme (Fig.3).

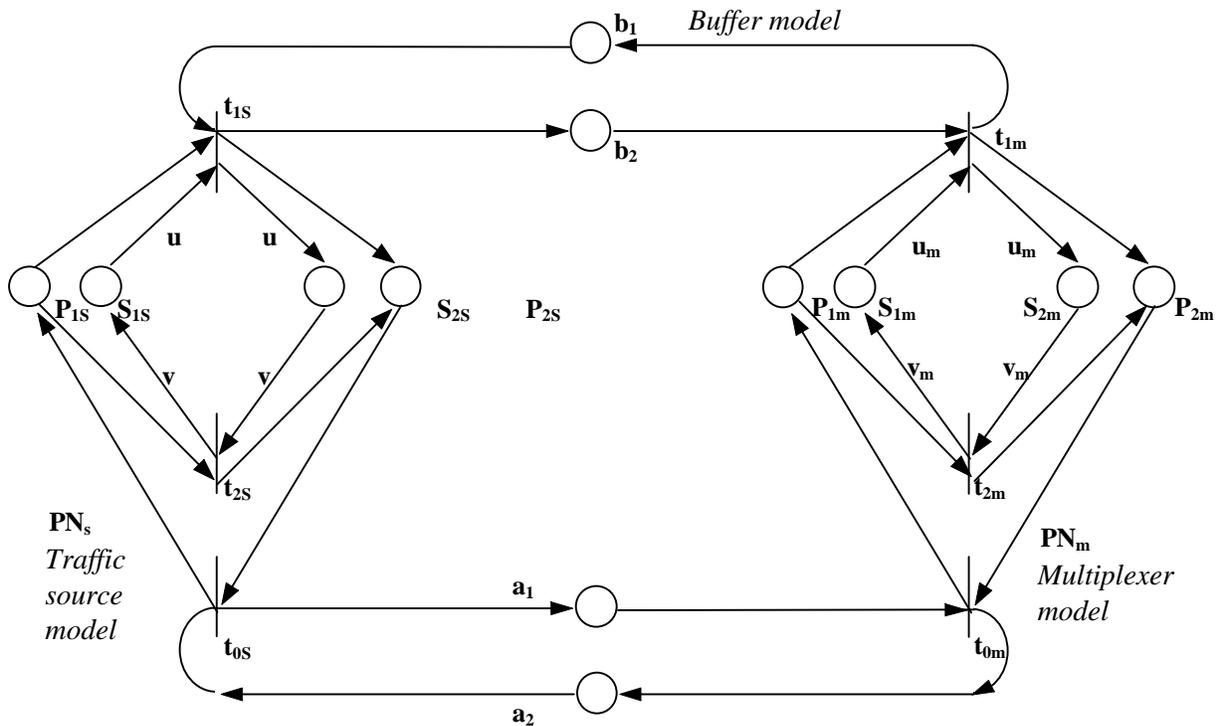

Fig.3

Let traffic of circuit $VC_i$ has parameters $PCR_i$, $SCR_i$, $MBS_i$. Such cell stream is simulated by firing sequence of transition $t_i$ of Petri net $PN_s$

*The model of object under control.*

Multiplexer work for every circuit can be represented as a flow of events - cell reads from appropriate buffer. Such flow can be modeled by Petri net $PN_m$ that has the same structure as $PN_s$. Since a virtual bandwidth $B_I$ for this channel is assigned on the basis of $SCR_i$

$$u_m=u, \quad v_m=v, \quad M(s_{1m})=u_m+ v_m-1.$$

The firing sequence of transition $t_{1m}$ will accord to guaranteed flow of cell reads from buffer.

In order to the models of traffic source and switch to work at the same time scale let us connect their "internal clocks" ($t_{0S}$ and $t_{om}$) via positions $a_1$ and $a_2$ and put one marker to position $a_2$. In that case the respective internal clocks operate at the same frequency but delayed in phase against each other.

Virtual buffer is simulated by positions $b_1$, $b_2$ that unite transition $t_{1s}$ (putting a cell to a buffer) and $t_{1m}$ (reading from a buffer). The number of tokens in $b_1$ and $b_2$ respectively, corresponds to the number of available and occupied buffer locations.

*The formal definition of efficiency criterion.*

The problem to be investigated is to determine the minimal buffer size when there is no cell loss caused by limited buffer capacity. In terms of Petri Net it corresponds to determining of the minimal number of tokens $M(b_1)$ in the place $b_1$ under which the transitions $t_{1S}$ and $t_{1m}$ may fire independently on $M(b_1)$. To solve this problem let us see the model with unlimited buffer, - an "open" net model from which the buffer positions $b_1$ and $b_2$ are removed so that the source model and multiplexer model execute independently from buffer size.

In this open model we determine the synchronic distance $syn(t_{1s}\ t_{1m})$.

The incidence matrix of this "open" net is:

|   | $t_{0S}$ | $t_{1S}$ | $t_{2S}$ | $t_{0m}$ | $t_{1m}$ | $t_{2m}$ |
|---|---|---|---|---|---|---|
| $S_{1S}$ |  | -u | v |  |  |  |
| $S_{2S}$ |  | u | -v |  |  |  |
| $P_{1S}$ | 1 | -1 | -1 |  |  |  |
| $P_{2S}$ | -1 | 1 | 1 |  |  |  |
| $a_1$ | 1 |  |  | -1 |  |  |
| $a_2$ | -1 |  |  | 1 |  |  |
| $S_{1m}$ |  |  |  |  | $-u_m$ | $v_m$ |
| $S_{2m}$ |  |  |  |  | $u_m$ | $-v_m$ |
| $P_{1S}$ |  |  |  | 1 | -1 | -1 |
| $P_{2S}$ |  |  |  | -1 | 1 | 1 |

I ==

From the set of linear equations

**I*h=0**

get reproduction vector

$r = (u+v,\ u, v,\ u_m+v_m,\ u_m,\ v_m)^T$,

Then the synchronic distance

$Syn(t_{1s}\ t_{1m}) = \max(|h_{1s}(p) - h_{1m}(p)|) = (M(s_{1S})/u)*(1 - v_m/(u_m+v_m))$

Since synchronic distance defines maximum possible difference between firing counts of $t_{1s}$ (putting a cell to a buffer) and $t_{1m}$ (taking a cell out from a buffer), the $Syn(t_{1s}\ t_{1m})$ equals the buffer capacity needed.

$S_i = M(b1) \geq Syn(t_{1s}\ t_{1m})$

Using the equations (1) we have

$S_i \geq MBS_i(1 - C_i/PCR_i)$

In the common case the virtual channel bandwidth assigned $C_i$ (guaranteed unloading rate) may be greater then $SCR_i$ and the buffer capacity needed will be less.

We assume that the value of **CDV** is mostly caused by the buffer unloading time $CDV = S_i/C_i$.

The efficiency criterion is

$J = m*S_i + C_1$ ,

where m is the weight coefficient.

Then optimization task is formulated as following linear programming task:

$J = m*S_i + C_1$ , $\longrightarrow$ Min ,

$S_i \geq MBS_i(1-C_i/PCR_i)$ ,

$S_i \leq CDV*C_i$ ,

$C_i \geq SCR_i$ ,

$\Sigma C_i \leq C$ ,

$\Sigma S_i \leq S$ ,

Let us note that models above allow to perform only the static distribution of network resources oriented to the worst case, as the traffic parameters taken describe the worst case restrictions of traffic contract and are not sufficient to define the traffic source behavior as a function of time and to synthesize an optimal real time control system. The additional information is needed to describe the stochastic nature and time dependency of a cell flow for the connection.

It is necessary to study different application traffics to determine their consistent time and distribution characteristics as well as to evaluate the adequacy of IND-model as a Marcovian realization.

*Traffic dynamics and traffic source model*

Despite all its outstanding capabilities, ATM technology has `some internal features that can cause traffic related problems. According to [12], even in lightly loaded cell-switching networks it is not unusual to observe cell blocking due to network congestion. The loss of a single cell in an ATM transmission leads to higher-level packet retransmission and an exponential increase in traffic through a switch. A congestion collapse is of particular concern when ATM switch relay large amounts of bursty data. In order to efficiently manage the network resources an additional research in traffic dynamics is needed.

Measurements of LAN and VBR video traffic show that data flow in packet and cell based networks differs fundamentally from conventional telephone traffic, and may be even fractal in nature [2]. In this case some conventional QoS criteria associated with actual bursty traffic may differ drastically from that predicted by conventional models. It should be taken into account while creating the MIX-model of cell flow. We offer to use the self-similar nature of traffic at synthesis of optimal control algorithm.

Below we show that the actual nature of packet traffic has serious implications for network engineering and management approach.

We will use a function of time **F(t)** as a model of input packet traffic. If the function is well-known the control decision is possible in off-line mode. Let **F(t)** is a stochastic function of time representing the rate of cell arriving. We assume that **F(t)** is a stationary in band sense stochastic process and we know its mean value **<F(t)>** and autocorrelation function $\mathbf{K_F(t)}$. Below we show that the use of autocorrelation function permits:

to get a closed form solution of the optimal traffic management task, and
to use the feature of traffic self-similarity to predict the traffic behavior in different time intervals.

Statistical analysis of high-resolution measurements of several types of network traffic has show that many type of network traffic are self-similar in nature[2]. That feature displays itself in fact that normalized correlation function COR/VAR of **F(t)** keeps its appearance in different observation intervals.

In Fig. 5 outcomes of measurement of a correlation function $\mathbf{K_F}(\tau)$ of the IP over ATM cell flow are shown for different time scales. The measurements were conducted in a network RUSnet [10] at use of FTP service.

The feature of traffic self-similarity permits to compute the correlation function for more short interval and use the same estimations for forecasting the traffic volume in order to control it for more long time.

*An example of optimal control synthesis.*
The common requirement for mathematical model is its adequatennes to the real system behavior. To accommodate bursty traffic, several approaches are evolving: dynamic buffering mechanism with per-VC queuing, intelligently organized packet/cell discard algorithms, and ABR control algorithm which based on Resource Management (RM) cells. This cells indicate the presence or absence of congestion and force end systems to change the amount of data being sent into network. From a point of view of the automatic control theory these managing gears can have the uniform formal description.

First of all, consider the simple cell discarding algorithm in the buffer of the cell switch. The buffer provides additional resource for a limited time interval. That resource is equivalent to temporal extending channel bandwidth to smooth the burstness of traffic. Cells from virtual circuits (VC) are scheduled in statistical multiplexing order. If any one VC sends a large burst of data then the service rate at the buffers of all other VC drops until the burst has been served. We will call the buffer for VC as a virtual buffer (VB). To protect the VB from congestion, a cell discarding scheme may be used as shown in Fig. 4.

The aims of the control are to maintain the number of cells in the VB queue at a desired setpoint, and protect buffer from overflow. Since the data flow has burst components, a border of the buffer's filling will oscillate around the setpoint value.

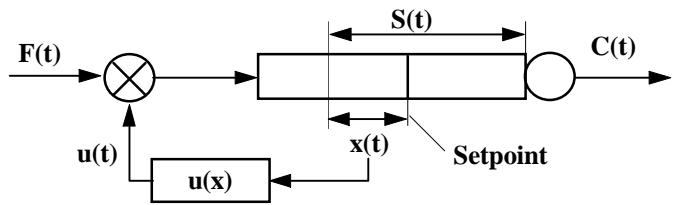

Fig. 4

The setpoint value depends on a assigned virtual bandwidth $\mathbf{C_i}$. The choosing of the $\mathbf{C_i}$ (at the SetUp phase of connection) reflects a tradeoff between mean cell delay, cell loss and bandwidth loss. It is determined on more high level of management (resources allocation level) and is transmitted to a level of cell discarding as parameter **C**.

*The model of input variables.*
As we have said above we will use the fluid approximation of data flow and use a stochastic process **F(t)** with mean value **<F(t)>** and autocorrelation function $K_F(t)$ as a model of input cell rate.

*The model of object under control.*
Let: **S(t)** - current buffer's filling, **<S(t)>** - mean value of **S(t)** (setpoint), **<F(t)>** - mean value of **F(t)**.

Then we ignore cells boundaries and define a speed of the buffer filling **dS/dt** by

$$dS/dt = F(t) + u(t) - C,$$

where **C** is the assigned (contract) service rate and **u(t)** - control rate for discarding cells. In this equation **S(t)** can be positive as well as negative. Negative values characterize a degree of not using of available VC bandwidth. Denote the value of a variable **x(t)** by

$$x(t) = S(t) - <S(t)>,$$

and a variable **f(t)** by

$$f(t) = F(t) - <F(t)>.$$

In case of mean value **<F(t)>** equal to service rate **C**, we can write the buffer equation in derivations

$$dx/dt = f(t) + u(t). \qquad (2)$$

*Formal definition of efficiency criterion.*
The efficiency criteria have to reflect the aim of control and take into account the values of QoS parameters needed by the particular traffic contract. In addition while choosing the criterion the designers have to minimize the complexity of the control value computation. In our case we have following functional

$$J = \frac{1}{T}\int_0^T Q(x(t))dx.$$

If $Q(x)=x^2$ then **J** denote as

$$J=<x^2>.$$

The requirements to minimization of CLR (Cell Loss Ratio) value can be expressed by means of integral limitation $<u^2> \leq N_u$ and efficiency criterion expressed as weighted sum

$$J = m^2<x^2> + <u^2>, \qquad (3)$$

where $m^2$ - weight coefficient. Its value depends on value of $N_u$.

The first item in **J** enforce the minimizing of Cell Delay Variation (CDV) the second item enforce the minimizing of Cell Loss Ratio (CLR). The choosing the value of $N_n$ reflects the relative significance of CLR and CDV regulating.

The depth of criterion minimum that can be reached depends on information of input data streams. In order to compute the minimum of **J** it is possible to use spectral density $A_f(\omega)$, that is Fourier transform of autocorrelation function $K_f(t)$.

The practical value of solving formulated optimization task depends on precision of defining of spectral density numerical characteristics.

The traffic feature of self-similarity can be used for getting those values.

The estimation $A_f(\omega)$ computed for more short interval can be used for forecasting the traffic volume for more long time. The result of experimental dependence approximation is shown at Fig.6.

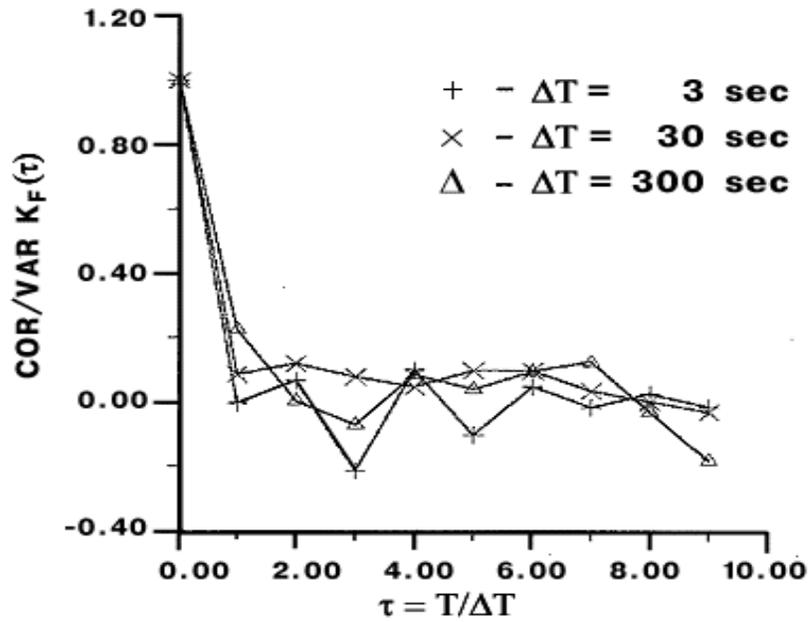

Fig. 5

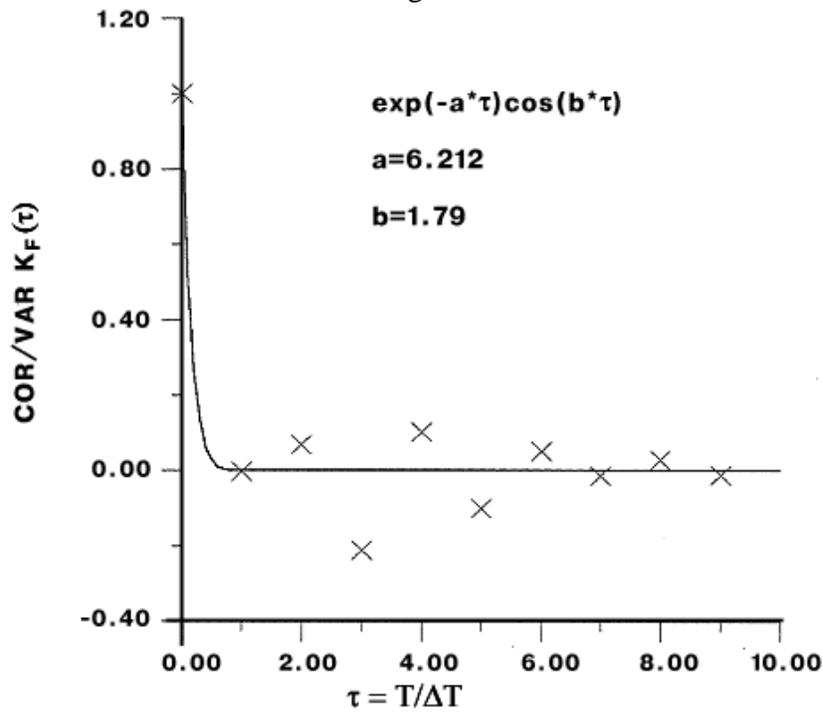

Fig.6

For approximation of a correlation function (Fig 6) we use the expression
$$K_f(\tau) = D_f \cdot e^{-a \cdot \tau} \cdot \cos(b\tau)$$, where **a, b** - parameters of approximation.
Then in case of $D_f = 1$ we can receive an explicit expression for spectral density $A_f(\omega)$

$$A_f(\omega) = \frac{2}{\pi}\int_0^\infty K_f(\tau)\cos(\omega\tau)d\tau = \frac{a}{\pi}\left[\frac{1}{a^2+(b-\omega)^2}+\frac{1}{a^2+(b+\omega)^2}\right]$$

Using the equations (2) and (3) we obtain expressions for spectral densities $A_u(\omega)$ and $A_x(\omega)$

$$A_u(\omega) = A_f(\omega) \cdot \frac{m^4}{(\omega^2+m^2)^2}; \qquad A_x(\omega) = A_f(\omega) \cdot \frac{\omega^2}{(\omega^2+m^2)^2},$$

Value **S** of $<u^2>$, $<x^2>$ and **J** could be calculated

$$\langle u^2\rangle_{min} = \varphi(m) = \frac{m}{2[(a+m)^2+b^2]}\left[(a+m)+m\frac{(a+m)^2-b^2}{(a+m)^2+b^2}\right],$$

$$\langle x^2\rangle_{min} = \psi(m) = \frac{1}{2m[(a+m)^2+b^2]}\left[(a+m)-m\frac{(a+m)^2-b^2}{(a+m)^2+b^2}\right],$$

$$J_{min}(m) = m^2\langle x^2\rangle_{min} + \langle u^2\rangle_{min} = \frac{m(a+m)}{(a+m)^2+b^2}.$$

This value of **J** is achieved by using a linear feedback law

$$u(t) = m^2\int_{t_0}^t x(\tau)d\tau - 2f(t_0) \ .$$

Where the latter item is received from the conditions of a transversality (boundary conditions) ensuring **x=0** in the beginning and end of an interval of regulation.

As we know value of $N_u$ and using the dependence $<u^2>_{min} = \varphi(m)$ shown in Fig. 7, it is possible to choose value of weight coefficient **m** and determine the feedback intensity **u(t)**.

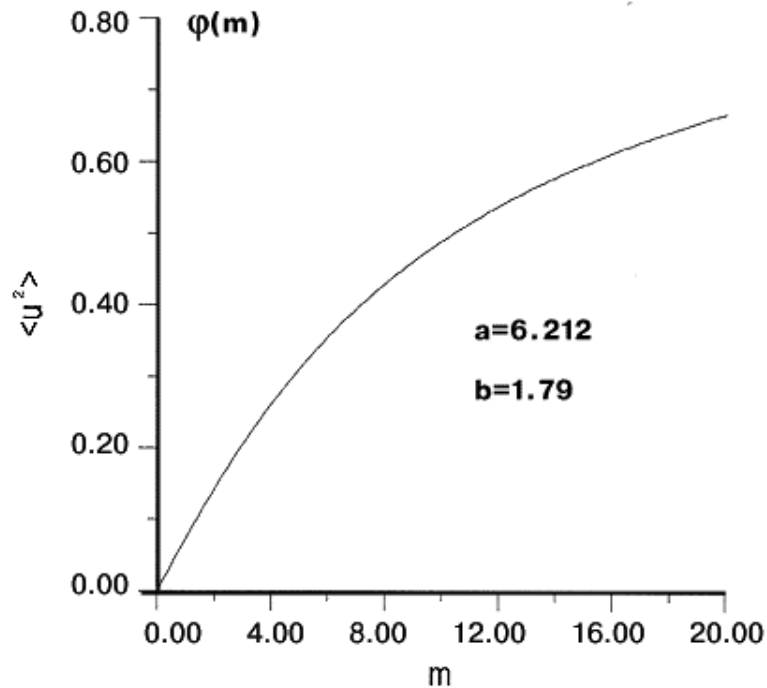

Fig. 7

Note, the received value of control action **u(t)** can be used in discarding cells algorithm as well as to change the source intensity in ABR mode. This circumstance has to be taken under consideration to synthesize the controls needed.

Relation **J=J(m)** in Fig. 8 allows to evaluate the numerical value of efficiency of the given control mechanism.

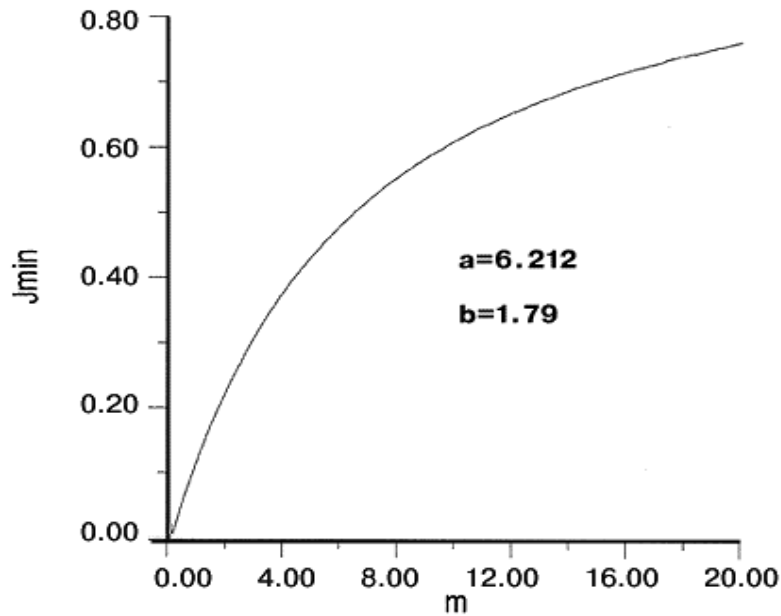

Fig. 8

The task of control optimization above can be generalized for multidimensional case .

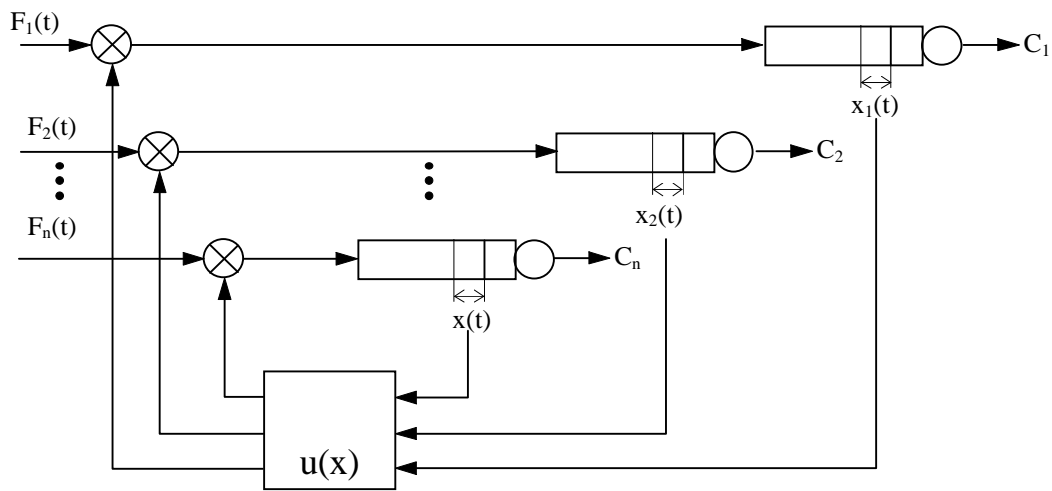

Fig. 9

Let us see that generalization for the example of dynamic bandwidth allocation scheme. At figure 7 value $C_i$ denotes value of sustainable cell rate (SCR) for **$VC_i$** that accords to set connection contract.

Taking into account the traffic dynamics permits to formulate the optimization task as two-step procedure. At the first step, the maximum value of possible increasing $VC_i$ bandwidth is being determined. To do that, it is possible, for example, formulate following linear programming task:

$$\begin{cases} \sum \gamma_i N_i \to \max_{\vec{N}} \\ N_i \leq \alpha_i \\ \sum N_i \leq \beta \end{cases} \quad \text{Where} \quad \begin{array}{l} N_i \text{ - } \textit{value to be found,} \\ \gamma_i \text{ - } \textit{weight coefficients,} \\ \alpha_i \text{ and } \beta \text{ - } \textit{parameters defined by the contract.} \end{array}$$

In our case the equation of dynamic processes is:

$$\frac{d\vec{x}}{dt} = \vec{f} + \vec{u}$$

and quality criterion is at the second step:

$$J = \frac{1}{T}\int_{t_0}^{T}\left(\vec{x}^T M \vec{x} + \vec{u}^T \vec{u}\right)dt \quad ,$$

where **M** - diagonal matrix with elements $M_{i,j} = \begin{cases} m_i^2, i = j \\ 0, i \neq j \end{cases}$

In that case the equation for control actions is:

$$M\vec{x} = \dot{\vec{u}}$$

under restrictions: $\begin{cases} t = t_0: \quad \vec{x} = 0, \quad \vec{u} = -2\vec{f}; \\ t = T: \quad \vec{x} = 0, \quad \vec{u} = 0 \end{cases}$

All weight coefficients in matrix **M** can be computed basing on valid values of $N_i$ gotten at the first step of optimization.

On Fig.10 interdependence of integral characteristics $<x^2>$ and $<u^2>$ is shown.

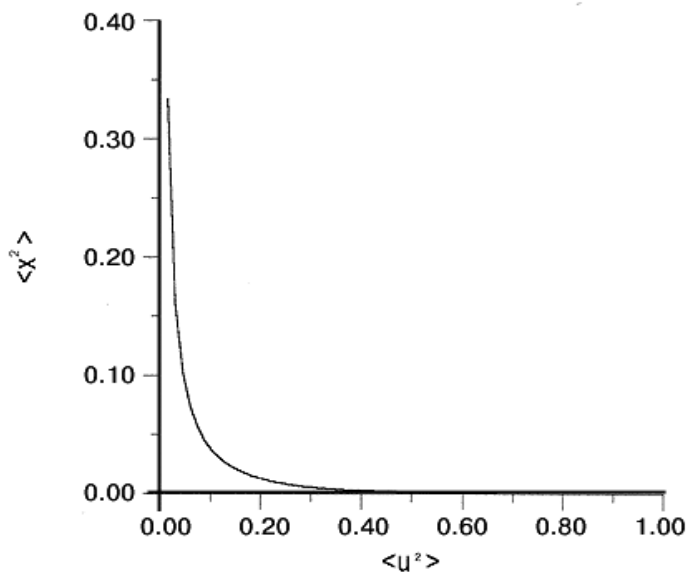

Fig. 10

*Conclusions.* Automatic control approach is proposed to formalize traffic management system design. The main steps of automatic control methodology include the hierarchical representation of management system and the formal definitions of input variables, object and goal of control of each management level. Requirements to the input variable models are discussed. We propose to distinguish two classes of input variables models. These models should be based on the actual individual traffic descriptors and on the real network workload parameters. A Petri net model of individual traffic source based on ATM forum specification is presented. It is noted that the current set of traffic parameters recommended by ATM-forum is not enough to synthesize optimal control system. It is necessary to explore each applications' traffic to determine its time and distribution characteristics. An example of an optimal control scheme for cell discarding algorithm is presented

The feature of traffic self-similarity can be used to effectively solve optimal control task. That feature displays itself in keeping the correlation function structure for different averaging times. The obtained estimations of the spectral density can be used for predicting and optimizing control for cell discarding algorithm and for dynamic bandwidth allocation mechanism.